# Document Searching System based on Natural Language Query Processing for Vietnam Open Courseware Library

**Dang Tuan NGUYEN and Ha Quy-Tinh LUONG**

**Faculty of Computer Science**

**University of Information Technology, Vietnam Nationnal University HCMC**

**Ho Chi Minh City, Vietnam**

## Abstract

The necessary of buiding the searching system being able to support users expressing their searching by natural language queries is very important and opens the researching direction with many potential. It combines the traditional methods of information retrieval and the researching of Question Answering (QA). In this paper, we introduce a searching system built by us for searching courses on the Vietnam OpenCourseWare Program (VOCW). It can be considered as the first tool to be able to perform the user's Vietnamese questions. The experiment results are rather good when we evaluate this system on the precision and the run-time of answering the Vietnamese questions.

**Keywords**: *Natural Language Processing, Document Retrieval, Search, Question Answering, Knowledge Base.*

## 1. Introduction

The necessary of buiding the searching system being able to support users expressing their searching by natural language queries is very important and opens the researching direction with many potential. It combines the traditional methods of information retrieval and the researching of Question Answering (QA).

Several searching engineering systems with supporting English language in e-library based on the natural language query processing were built by our previous publications [1], [2], [3], [4], [5], [6], [7], [8]. In the continued researches, we aimed to developing a document searching system base on answering to the user's Vietnamese questions.

In this paper, we introduce a searching system built by us for searching courses on the Vietnam OpenCourseWare Program[1] (VOCW). It can be considered as the first tool to be able to perform the user's Vietnamese questions. The experiment results are rather good when we evaluate this system on the precision and the run-time of answering the Vietnamese questions.

## 2. System Architecture

The system model is represented in the Fig. 1, including the main components as follow:
- The extracting component: extracting the data in the metadata pages of VOCW.
- The ontology-based knowledge (VOCW ontology) for storing VOCW data.
- The parsing component: analyzing the syntactic of Vietnamese question and returning the parse tree and generating tree. The generating tree will be the input of the generating query. It operates on the set of the defined syntactic rules for analyzing Vietnamese question.
- The generating component: generating SPARQL[17] query for the querying component.
- The querying component: inserting, updateting and querying them to the VOCW ontology.

On this system model, the system processes as follow:
- When the user inputs the Vietnamese question, the parsing component will analyze its syntax. We defined the set of EBNF syntactic rules before. If user's question syntactic isn't belong to this set, the system returns that this question isn't analyzed (no syntactic exactly) and then stop program. By contrast, it returns the parser tree and the generating tree. The generating component will use the generating tree for creating SPARQL query. And the querying component gets the data on VOCW ontology.
- The extracting component extracts the data in the metadata pages of VOCW. And the querying component inserts the extracted data to VOCW ontology.

[1] http://vocw.edu.vn/

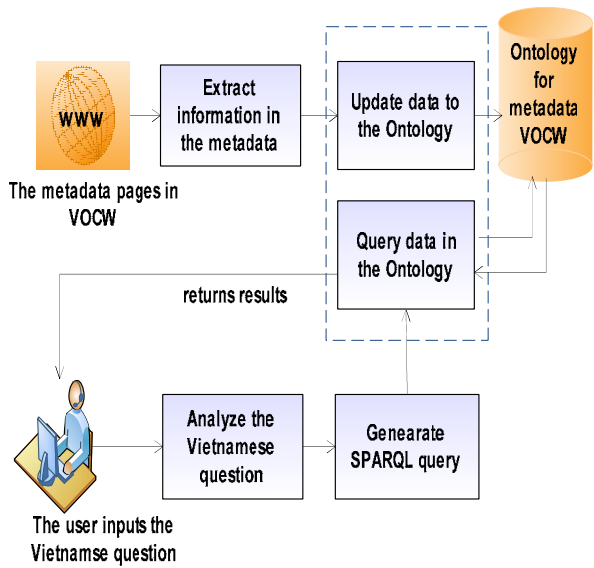

The metadata pages in VOCW

returns results

The user inputs the Vietnamse question

Fig. 1 System architecture

For example, we consider the Vietnamese question:

- *"Ai đã viết sách Toan?"*
  *("Who wrote Toan book?")*

This question will be syntactically parsed and then will output the parse tree and generating tree as follow:

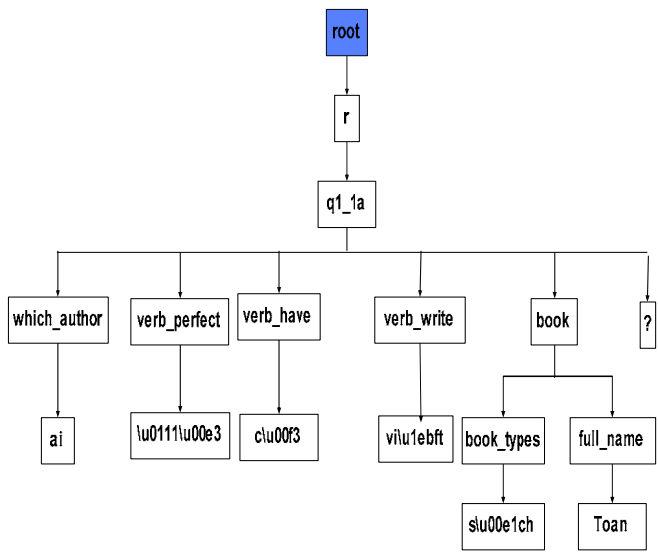

Fig. 2 Syntactic tree of a Vietnamese query (base on antlrworks generated parse tree)

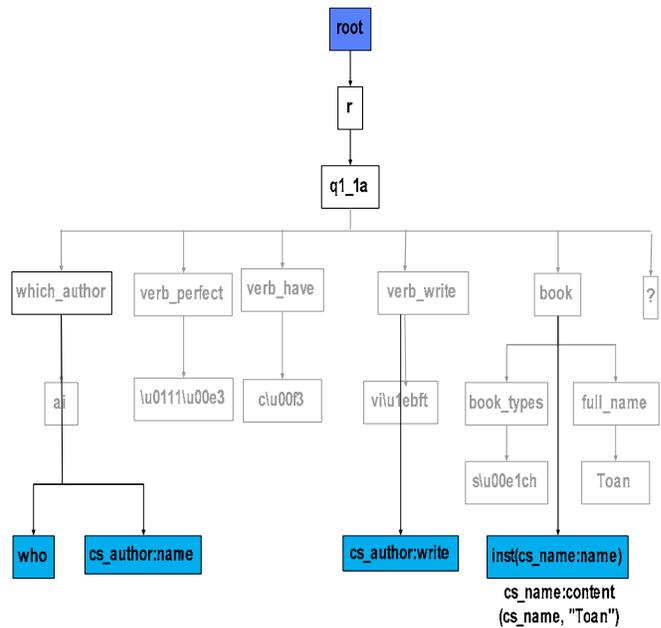

Fig. 3 The generating tree

The SPARQL query is created from the generating tree:

```
SELECT DISTINCT ?authorname
FROM <http://localhost/owl_test/vocw_full.owl>
WHERE {
    {?author cs_author:content ?authorname}.
    {?author cs_author:write ?course}.
    {?year cs_created:content ?yearname
    FILTER regex(?yearname , "^2009$", "i" )}.
    {?course cs_name:isWrittenIn ?year}.
    {?course cs_name:content ?coursename
    FILTER regex(?coursename , "^cấu trúc dữ liệu$", "i" ).
    }
}
```

## 3. Syntactic Rules

The parsing component is built for analyzing Vietnamese questions regarding to the data such as name, language, summary, authors, copyright holders, maintainers keywords, version and affiliations.

For examples, some different questions as follow:

- *Ai đã viết cuốn sách B vào năm 2000?*
  *(Who wrote book B in 2000?)*
- *Nhà xuất bản nào đã phát hành cuốn B trong năm 2008?*
  *(Which publisher published book B in 2008?)*
- *Sách B được tác giả A viết vào năm nào?*
  *(What year did writer A write book B?)*

Syntax of Vietnamese questions can be built by EBNF

notation (Extended Backus–Naur Form). We defined 40 syntactic rules for Vietnamese questions regarding the above data.

In this part, we use ANTLR tool for generating lexer and parser code in order to analyzing and checking syntax of Vietnamese questions (set of the defined EBNF syntactic rules). And we design this component for adding new rules easily but no modifying code much in future.

TABLE 1. SYNTACTIC RULES

| No | Syntactic rules |
|----|-----------------|
| 1 | <Q1.1a> = <what_author> [<vperfect>] [<interrogative1>] <verb_write> <book> {[<conjunction>] <book>} [<time_phrase>] "?" |
| 2 | <Q1.1b> = [<time_phrase>] ["","] <what_author> [<vperfect>] [<interrogative1>] <verb_write> <book> {[<conjunction>] <book>} "?" |
| 3 | <Q1.1c> = <book> {[<conjunction>] <book>} [<vperfect>] <vpassive> <what_author> <verb_write> [<time_phrase>] "?" |
| 4 | <Q1.1d> = [<time_phrase>] ["","] <book> {[<conjunction>] <book>} [<vperfect>] <vpassive> <what_author> <verb_write> "?" |
| 5 | <Q1.2a> = [<interrogative3>] <creator> [<possessive>] <book> {[<conjunction>] <book>} <verb_be> <author> [<interrogative2>] "?" |
| 6 | <Q1.2b> = [<interrogative3>] <author> <verb_be> <creator> [<possessive>] <book> {[<conjunction>] <book>} [<interrogative2>] "?" |
| 7 | <Q1.2c> = <author> [<interrogative3>] <verb_be> <creator> [<possessive>] <book> {[<conjunction>] <book>} [<interrogative2>] "?" |
| 8 | <Q1.3a> = [<interrogative3>] <author> [<vperfect>] [<interrogative1>] <verb_write> <book> {[<conjunction>] <book>} [<time_phrase>] [<interrogative2>] "?" |
| 9 | <Q1.3b> = [<time_phrase>] ["","] [<interrogative3>] <book> {[<conjunction>] <book>} [<vperfect>] <vpassive> <author> <verb_write> [<interrrogative2>]"?" |
| 10 | <Q1.4a> = <author> [<vperfect>] [<interrogative1>] <verb_write> <book> {[<conjunction>] <book>} [<prep_time>] <what_time> "?" |
| 11 | <Q1.4b> ::= <book> {[<conjunction>] <book>} [<vperfect>] <vpassive> <author> <verb_write> [<prep_time>] <what_time> "?" |
| 12 | <Q2.1a> = <what_publisher> [<vperfect>] [<interrogative1>] <verb_publish> <book> {[<conjunction>] <book>} [<time_phrase>] "?" |
| 13 | <Q2.1b> = [<time_phrase>] ["","] <what_publisher> [<vperfect>] [<interrogative1>] <verb_publish> <book> {[<conjunction>] <book>} "?" |
| 14 | <Q2.1c> = <book> {[<conjunction>] <book>} [<vperfect>] <vpassive> <what_publisher> <verb_publish> [<time_phrase>] "?" |
| 15 | <Q2.1d> = [<time_phrase>] ["","] <book> {[<conjunction>] <book>} [<vperfect>] <vpassive> <what_publisher> <verb_publish> "?" |
| 16 | <Q2.2a> = [<interrogative3>] <publisher> [<vperfect>] [<interrogative1>] <verb_publish> <book> {[<conjunction>] |
| 17 | <book>} [<time_phrase>] [<interrogative2>] "?" |
| 17 | <Q2.2b> = [<time_phrase>] ["","] [<interrogative3>] <publisher> [<vperfect>] [<interrogative1>] <verb_publish> <book> {[<conjunction>] <book>} [<interrogative2>] "?" |
| 18 | <Q2.2c> = [<interrogative3>] <book> {[<conjunction>] <book>} [<vperfect>] <vpassive> <publisher> <verb_publish> [<time_phrase>] [<interrogative2>] "?" |
| 19 | <Q2.2d> = [<time_phrase>] ["","] [<interrogative3>] <book> {[<conjunction>] <book>} [<vperfect>] <vpassive> <publisher> <verb_publish> [<interrogative2>] "?" |
| 20 | <Q2.3a> = <publisher> [<vperfect>] [<interrogative1>] <verb_publish> <book> {[<conjunction>] <book>} [<prep_time>] <what_time> "?" |
| 21 | <Q2.3b> = [<prep_time>] <what_time> <publisher> [<vperfect>] [<interrogative1>] <verb_publish> <book> {[<conjunction>] <book>} "?" |
| 22 | <Q2.3c> = <book> {[<conjunction>] <book>} [<vperfect>] <vpassive> <publisher> <verb_publish> [<prep_time>] <what_time> "?" |
| 23 | <Q2.3d> = [<prep_time>] <what_time> <book> {[<conjunction>] <book>} [<vperfect>] <vpassive> <publisher> <verb_publish> "?" |
| 24 | <Q3.1a> = <book> [<of_author>][<by_publisher>][<time_phrase>] <is_of> <what_subject> ? |
| 25 | <Q3.1b> = [<time_phrase>] [,] <book> [<of_author>] [<by_publisher>] <is_of> <what_subject> ? |
| 26 | <Q3.1c> = [<field> <possessive>] <book> [<of_author>] [<by_publisher>] [<time_phrase>] <interrogative4> ? |
| 27 | <Q3.1d> = [<time_phrase>] [,] <field> <possessive> <book> [<of_author>] [<by_publisher>] <interrogative4> ? |
| 28 | <Q3.2a> = <book> [<of_author>] [<by_publisher>] [<time_phrase>] [<interrogative1>] <is_of> <subject> [<interrogative2>] ? |
| 29 | <Q3.2b> = [<time_phrase>] [,] <book> [<of_author>] [<by_publisher>] [<interrogative1>] <is_of> <subject> [<interrogative2>] ? |
| 30 | <Q3.2c> = <book> [<of_author>] [<by_publisher>] [<time_phrase>] [<interrogative3>] <verb_be> <book_type> <is_of> <subject> [<interrogative2>] ? |
| 31 | <Q3.2d> = [<time_phrase>] [,] <book> [<of_author>] [<by_publisher>] [<interrogative3>] <verb_be> <book_type> <is_of> <subject> [<interrogative2>] ? |
| 32 | <Q3.3a> = [<time_phrase>] [,] <author> [<vperfect>] [interrogative1] <verb_write> [<plural>] <book_type> <verb_have> <what_subject> ? |
| 33 | <Q3.3b> = <author> [<vperfect>] [interrogative1] <verb_write> [<plural>] <book_type> <verb_have> <what_subject> [<time_phrase>]? |
| 34 | <Q3.3c> = [<time_phrase>] [,] <book> [<of_author>] [interrogative1] <verb_write> [<plural>] <book_type> <is_of> <what_subject> ? |
| 35 | <Q3.3d> = <author> [<vperfect>] [interrogative1] <verb_write> [<plural>] <book_type> <is_of> <what_subject> |
| | [<time_phrase>] ? |
| 36 | <Q3.4a> = <publisher> [<vperfect>] [<interrogative1>] <verb_publish> [<plural>] <verb_have> <what_subject> [<time_phrase>] ? |
| 37 | <Q3.4b> = [<time_phrase>] <publisher> [<vperfect>] [<interrogative1>] <verb_publish> [<plural>] <verb_have> <what_subject> ? |
| 38 | <Q3.4c> = <publisher> [<vperfect>] [<interrogative1>] <verb_publish> [<plural>] <is_of> <what_subject> [<time_phrase>] ? |
| 39 | <Q3.4d> = [<time_phrase>] <publisher> [<vperfect>] [<interrogative1>] <verb_publish> [<plural>] <is_of> <what_subject> ? |
| 40 | <Q4.1a> = [plural] [book_type] [<verb_have> <subject>] [<by_author>] [<time_phrase>] <interrogative4> ? |
| 41 | <Q4.1b> = [<time_phrase>] [,] [plural][book_type] [<verb_have><subject>] [<by_author>] [interrogative4] ? |
| 42 | <Q4.1c> = [plural][book_type] [<is_of><subject>] [<by_author>] [<time_phrase>]<interrogative4> ? |
| 43 | <Q4.1d> = [<time_phrase>] [,] [plural][book_type] [<is_of><subject>] [<by_author>] <interrogative4> ? |
| 44 | <Q4.2a> = [plural] <book_type> [<verb_have> <subject>] by_publisher> [<time_phrase>] <interrogative4> ? |
| 45 | <Q4.2b> = [<time_phrase>][,][plural]<book_type> [<verb_have> <subject>] <by_publisher> <interrogative4> ? |
| 46 | <Q4.2c> = [plural]<book_type> [<is_of><subject>] <by_publisher> [<time_phrase>] <interrogative4> ? |
| 47 | <Q4.2d> = [<time_phrase>] [,] [plural] <book_type> <is_of> <subject> <by_publisher> <interrogative4> ? |
| 48 | <Q5.1a> = <book> [<vperfect>] <vpassive> [<publisher> <verb_publish> <what_place> [<time_phrase>] "?" |
| 49 | <Q5.1b> = [<time_phrase>] [","] <book> [<vperfect>] <vpassive> <verb_publish> <what_place> "?" |
| 50 | <Q5.2> = <publisher><verb_locate><what_place> "?" |
| 51 | <Q6.1a> = [<verb_buy>] <book> <verb_cost> "?" |
| 52 | <Q6.1b> = <price> [<possessive>] <book> [<what_price>] "?" |
| 53 | <Q7.1> = <how_many> <book> <in_elib> "?" |
| 54 | <Q7.2a> = <author> [<vperfect>] [<interrogative1>] <verb_write> <how_many> <book> [<time_phrase>] "?" |
| 55 | <Q7.2b> = [<time_phrase>] [","] <author> [<vperfect>] [<interrogative1>] <verb_write> <how_many> <book> "?" |
| 56 | <Q7.3a> = <publisher> [<vperfect>] [<interrogative1>] <verb_publish> <how_many> <book> [<time_phrase>] "?" |
| 57 | <Q7.3b> = [<time_phrase>] [","] <publisher> [<vperfect>] [<interrogative1>] <verb_publish> <how_many> <book> "?" |

# 4. Ontology Model

Ontology is built from the common terminologies containing concepts (terms), properties, definitions of its and relationships of its, including constraints. It supplies the possibility of the semantic representation and reasoning support

A statement is a triple (resource-property-value) for defining the property of the resource. "Resource" is considered as an object, a thing we want to speak. Example, Resource: authors, books, publishers, places, people, hotels…, URL, URI. "Property" is a kind of the particular resource, describing relations between resouces, for example, "written by", "age", "title"… And "value" can be a reaource or literal

Example:
- *Tác giả A viết sách B*
  *(Author A writes book B)*
is represented as follow:

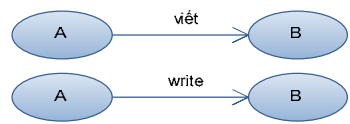

Fig. 4 Triple of write(cs_author,cs_name)

First, we build the semantic model between the object classes. Then, each statement is a triple, where resource and value are instances of classes.

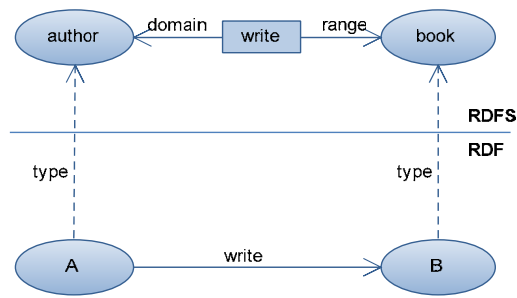

Fig. 5 The RDF/RDFS layers of write(cs_author,cs_name)

- *Resource: "Tác giả A" (Writer A)*
- *Property: "viết bởi" (written by)*
- *Value: : "sách B" (book B)*

In case of many statements: the example of the semantic net representing relations between "Writer", "Book", "Year" and "Publisher":

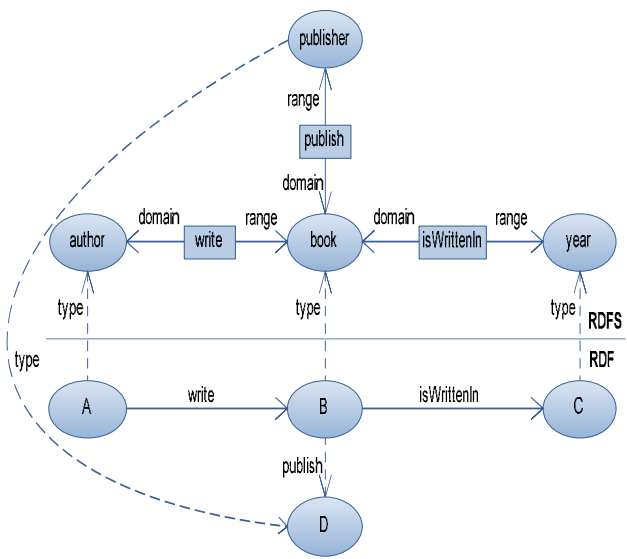

Fig. 6 The RDF/RDFS layers for "A write B book" and "B book is written by A" sentences

Terminology (resource, property, value) is organized, built and defined in hierarchy favourable to defining semantics exactly.

In the model above, we could add new properties from the existed properties (SWRL[14,15])

write(author,book) → isWrittenBy(book,author)

The vocwqa component carries out adding new data by SPARQL query language. Then, it checks the consitence by Pellet [16] và does inferences by Jess [15]

# 5. Generating Ontology Query

We consider single Vietnamese questions. It could contain one or many conjunctions "và" (and) or "hoặc" (or).
- In case of word "hoặc" (or): SPARQL supports keyword UNION for querying one possibility or more.
- In case of word "và" (and):

Example:
- *Ai đã có viết sách "Toan" và sách "Van" trong năm 2009?*

This sentence is understood in other way as follow:
*{Ai đã có viết sách "Toan" trong năm 2009 mà*
*{người này cũng viết sách "Van" trong năm 2009}*
*?}*

The example form:

*Who Author  Name[Toan,Van] ?*

We have:
- the resources and property: Author, Name.
- the question word: who.
- the value: Toan, Van.

This case is resolved with the nested SPARQL form. Because each node of the mapping tree could be root node of other subtree, the mapping syntactic tree could represent nested query in order to participate in SPARQL generating process.

We see that this case has the general form as follow:
- Set the resources, property and question word: $A_i$.
- Set the value: $x_i$.

The general form:
$$A_1 A_2 \ldots A_i[x_{i1}, x_{i2}, \ldots, x_{im}] \ldots A_k[x_{k1}] \ldots A_n?$$

Each question has only a component of word (terminology) questioned with a lot of the specific data by the linking-word "và" (and) or "hoặc" (or). Other components could be questioned at most a specific datum

Return example:
*{Ai đã có viết sách "Toan" mà*        (1)
*{người này cũng viết sách "Van" }*        (2)
*?}*

Its form:
*Who Author  Name[Toan]*
*[Who Author  Name[Van]] ?*

We see that sub-sentence (2) differ only sentence (1) in book "Van".

Let consider:
$$A_1 A_2 \ldots A_i[x_{i1}, x_{i2}, \ldots, x_{im}] \ldots A_k[x_{k1}] \ldots A_n?$$

We could analyze into m sentences. Sentences differ each other in component i.
$$A_1 A_2 \ldots Ax_{i1} \ldots A_i[xk_1] \ldots A_n?$$
$$A_1 A_2 \ldots Ax_{i1} \ldots A_i[xk_2] \ldots A_n?$$
…
$$A_1 A_2 \ldots Ax_{i1} \ldots A_i[xk_m] \ldots A_n?$$

On the syntactic tree, each node $A_i[xk_j]$ contains subtree. This subtree has root node $A_i[x_{kj}]$ including all subnode of parent-tree exception nodes $A_i[x_{ki}]_{(i=1..m)}$.

The above example will be generated to SPARQL query:

```
SELECT DISTINCT ?authorname
FROM <http://localhost/owl_test/vocw_full.owl>
WHERE {
    {?author cs_author:content ?authorname}.
    {?author cs_author:write ?course}.
    {?course cs_name:content ?coursename
    FILTER regex(?coursename , "^Toan$", "i" ).
    }
  .{
    SELECT DISTINCT ?authorname
    FROM <http://localhost/owl_test/vocw_full.owl>
    WHERE {
    {?author cs_author:content ?authorname}.
    {?author cs_author:write ?course}.
    {?course cs_name:content ?coursename
    FILTER regex(?coursename , "^Van$", "i" ).
    }}
  }
}
```

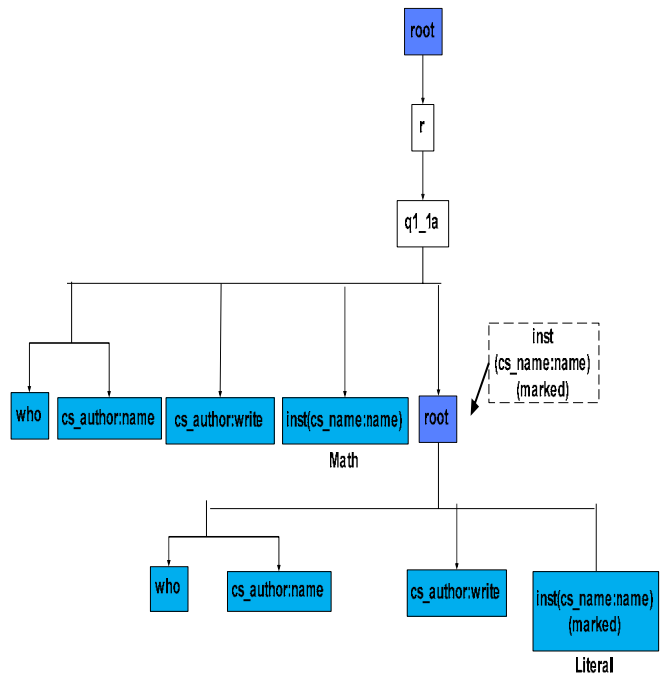

Fig. 7 The generating tree of "Who wrote Math and Literal book?"

## 6. Experiments

Base on the model system in Fig. 1, we developed the searching system with the specific characteristics for VOCW. It is evaluated according to the testing method that the track QA@INEX (Question Answering of INEX forum) used. We created the list of experiment questions and evaluated them manually by checking the answer appropriating to the question or not. According to this method, we didn't consider the satisfaction of the answer for the question. Thus, we only consider the "acceptability" of the answer for the question.

We developed the standard test suite including 40 Vietnamese questions corresponding with 40 syntactic rules, respectively. In addition, we collected more 91 questions (the random test collection) from users. We ran them on the computer with the configuration:
- Intel(R) Core(TM)2 Duo CPU, T5750 @ 2.00GHz, 997MHz, 0.99GB of RAM.
- WindowXP SP2

And we have the results as follow:
- The standard test suite:
    Percent test passed: 40/40 (100%)
- The random test collection:
    Percent test passed: 77/91 (84.62% test passed)
- The total run time: 2.54m /91
- The average run time of one question : 1.65s

With the above results and evaluation methods, the system is rather well.

## 7. Conclusions

We built the model, deployed the system for VOCW in reality. It is experimented and evaluated. And the results are rather good as the above observation. However, there are the first results in this application domain. The processing on the complex Vietnamese is a big challenging, so no real system in nowadays can solve it.

We concentrate on the research on the stronger syntactic parser for Vietnamese. The most important we need to extend the semantic model for the query and the methods and technologies for processing the semantic model in order to understanding by the computer.

The experience on the built system is useful for us to develop the other similar searching systems processing the user's Vietnamese question

## References

[1]  Dang Tuan Nguyen, Tuyen Thi-Thanh Do, "E-Library Searching by Natural Language Question-Answering System", Proceedings of the Fifth International Conference on Information Technology in Education and Training (IT@EDU2008), pages: 71-76, Ho Chi Minh and Vung Tau, Vietnam, December 15-16, 2008.

[2]  Dang Tuan Nguyen, Tuyen Thi-Thanh Do, "e-Document Retrieval by Question Answering System", International Conference on Communication Technology, Penang, Malaysia, February 25-27,

2009. Proceedings of World Academy of Science, Engineering and Technology, Volume 38, 2009, pages: 395-398, ISSN: 2070-3740.

[3]    Dang Tuan Nguyen, Tuyen Thi-Thanh Do, "Natural Language Question Answering Model Applied To Document Retrieval System", International Conference on Computer Science and Technology, Hongkong, China, March 23-25, 2009. Proceedings of World Academy of Science, Engineering and Technology, Volume 39, 2009, pages: 36-39, ISBN: 2070-3740.

[4]    Dang Tuan Nguyen, Tuyen Thi-Thanh Do, "Document Retrieval Based on Question Answering System", Proceedings of the Second International Conference on Information and Computing Science, pages: 183-186, Manchester, UK, May 21-22, 2009. ISBN: 978-0-7695-3634-7. Editions IEEE.

[5]    Dang Tuan Nguyen, Tuyen Thi-Thanh Do, Quoc Tan Phan, "A Document Retrieval Model Based-on Natural Language Queries Processing", Proceedings of the International Conference on Artificial Intelligence and Pattern Recognition (AIPR), pages: 216-220, Orlando, FL, USA, July 13-16, 2009. ISBN: 978-1-60651-007-0. Editions ISRST.

[6]    Dang Tuan Nguyen, "Interactive Document Retrieval System Based-on Natural Language Query Processing", Proceedings of the Eighth International Conference on Machine Learning and Cybernetics, pages: 2233-2237, Baoding, Hebei, China, July 12-15 2009. ISBN: 978-1-4244-3703-0. Editions IEEE.

[7]    Dang Tuan Nguyen, Tuyen Thi-Thanh Do, Quoc Tan Phan, "Integrating Natural Language Query Processing and Database Search Engine", Proceedings of the 2009 International Conference on Artificialal Intelligence - ICAI'09, Volume 1, pages: 137-141, Las Vegas, Nevada, USA, July 13-16, 2009. ISBN: 1-60132-107-4, 1-60132-108-2 (1-60132-109-0). CSREA Press.

[8]    Dang Tuan Nguyen, Tuyen Thi-Thanh Do, Quoc Tan Phan, "Natural Language Interaction-Based Document Retrieval", *The 2nd IEEE International Conference on Computer Science and Information Technology 2009* (ICCSIT 2009), Volume 4, pages: 544-548. Beijing, China, August 8-11, 2009. ISBN: 978-1-4244-4520-2. Editions IEEE.

[9]    Dragan Gas˘evic´, Dragan Djuric´, Vladan Devedz˘ic' (2006), Model Driven Architecture and Ontology Development, page 58-68, 91-108, p.107-108.

[10]   Grigoris Antoniou và Frank van Harmelen (2004), A Semantic Web Primer, The MIT Press Cambridge, Massachusett, London, England, p.31-33.

[11]   Natalya F. Noy và Deborah L. McGuinness (2001), Ontology Development 101: A Guide to Creating Your First Ontology

[12]   Terence Parr, The Pragmatic Bookshelf, Raleigh, North Carolina Dallas, Texas (2007), The Definitive ANTLR Reference, Building Domain-Specific Languages,  page 98-99

[13]   Franz Baader, Deborah L. McGuinness, Daniele Nardi, Peter F. Patel-Schneider (), THE DESCRIPTION LOGIC HANDBOOK: Theory, implementation, and applications, page 47-100, page 495-505

[14]   Protégé (http://protege.stanford.edu/)

[15]   Jess (http://www.jessrules.com/)

[16]   Pellet (http://clarkparsia.com/pellet/)

[17]   Jena (http://jena.sourceforge.net/)